\documentclass[twocolumn]{aastex631}

\usepackage{algorithm}
\usepackage{algpseudocode}

\newcommand{\code}[1]{\texttt{#1}}
\newcommand{\mesa}{\code{MESA}}
\newcommand{\MESA}{\mesa}
\newcommand{\mesatest}{\code{mesa\_test}}

\submitjournal{ApJS}
\shorttitle{Testing MESA}
\shortauthors{Wolf et al.}
\received{August 29, 2023}
\revised{September 28, 2023}
\accepted{October 04, 2023}

\begin{document}

\title{Testing Modules for Experiments in Stellar Astrophysics (MESA)}

\correspondingauthor{William M. Wolf}
\email{wolfwm@uwec.edu}

\author[0000-0002-6828-0630]{William M. Wolf}
\affiliation{Department of Physics and Astronomy, University of Wisconsin, Eau Claire, WI 54701, USA}

\author[0000-0002-4870-8855]{Josiah Schwab}
\affiliation{Department of Astronomy and Astrophysics, University of California, Santa Cruz, CA 95064, USA}

\author[0000-0003-3441-7624]{R.~Farmer}
\affiliation{Max-Planck-Institut f\"{u}r Astrophysik, Karl-Schwarzschild-Straße 1, 85741 Garching, Germany}

\author[0000-0002-4791-6724]{Evan B. Bauer}
\affiliation{Center for Astrophysics $\vert$ Harvard \& Smithsonian, 60 Garden St, Cambridge, MA 02138, USA}

\begin{abstract}
  Regular, automated testing is a foundational principle of modern
  software development.  Numerous widely-used continuous integration
  systems exist, but they are often not suitable for the unique needs
  of scientific simulation software.  Here we describe the testing
  infrastructure developed for and used by the Modules for Experiments
  in Stellar Astrophysics (MESA) project.  This system allows the
  computationally-demanding MESA test suite to be regularly run on a
  heterogeneous set of computers and aggregates and displays the
  testing results in a form that allows for the rapid identification
  and diagnosis of regressions.  Regularly collecting comprehensive
  testing data also enables longitudinal studies of the performance of
  the software and the properties of the models it generates.
\end{abstract}


\section{Introduction}

The stellar evolution software instrument \MESA\ \citep[Modules for Experiments in Stellar Astrophysics;][]{Paxton2011,Paxton2013,Paxton2015,Paxton2018,Paxton2019,Jermyn2023} is developed by a worldwide team.
The complex nature of the software itself and of the stellar evolution
scenarios that it models demand frequent, automated testing in order
to ensure changes do not have unintended side effects.

To this end,
\MESA\ has a collection of approximately 100 test cases spanning a
wide range of stellar modeling applications that are regularly run and
checked for regressions.
These test cases can take anywhere from tens of seconds to many hours to
complete, varying with the particular test case and the machine running it.
\MESA{} has a large user base that runs it on a heterogeneous set of Unix-based
operating systems and computers ranging from consumer laptops up to
supercomputing clusters. Additionally, the global nature of \MESA{} development
requires that results from running the test suite need to be easily accessible
through a cloud-based interface. These three aspects (lengthy test cases,
heterogenous hardware and software, and distributed development) require a unique
testing infrastructure, which we lay out in this paper. 

Aspects of this testing infrastructure, including the \MESA{} TestHub, were
briefly introduced in Appendix D of \citet{Paxton2019} as a means to aggregate
and digest test results on a variety of different hardware and software
platforms.
Since that time, the \MESA\ testing infrastructure has been
extensively overhauled to reflect changes to the \MESA{} development
workflow (i.e., a change of version control system from SVN to Git)
and to expand its capabilities.
Several of these updates were briefly described in \citet{Jermyn2023}. This
article gives a more complete summary of the current design and implementation
of the \MESA\ testing infrastructure. We hope that other research software
projects with similar needs will use our findings to support their own testing
infrastructures.
\section{Testing}

\begin{figure*}
  \includegraphics[width=\textwidth]{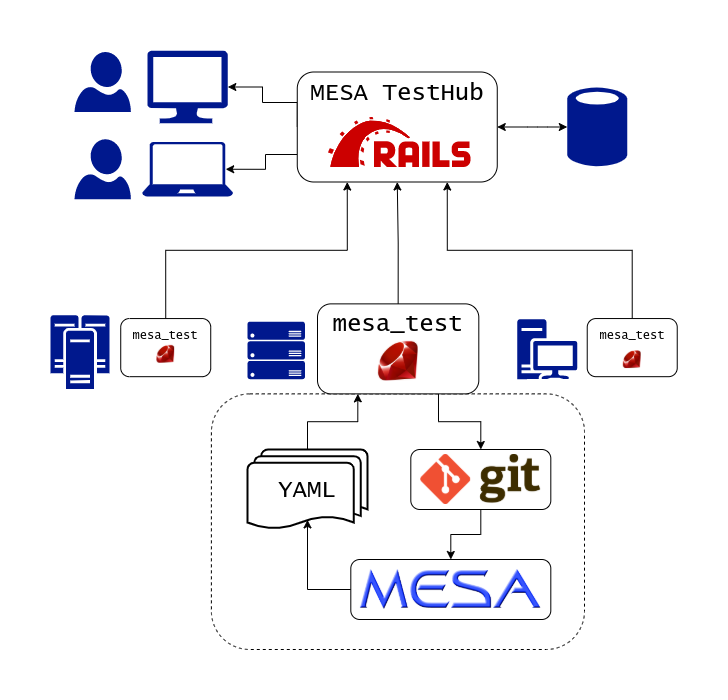}
  \caption{Workflow for testing \MESA{}.  Developers interact with the \MESA{} TestHub via a web browser.  The data for the TestHub comes from instances of \mesatest\ run on a heterogeneous set of computer systems (laptops, desktops, clusters).  The internal operation of one instance of \mesatest\ is expanded and illustrated in the dotted rectangle. \label{fig:flowchart}}
\end{figure*}

Figure~\ref{fig:flowchart} summarizes the \MESA\ testing flow.  At the
lowest level shown, \MESA\ itself contains scripts that compile the
software and execute tests (Section~\ref{sec:mesa}).  The process of
obtaining a specific version of \MESA\ and invoking these scripts is
managed by the \mesatest\ Ruby gem (Section~\ref{sec:mesa_test}),
which also transmits the results of the tests.  The
data reported from instances of \mesatest\ are collected and
collated by the \MESA{} TestHub Ruby on Rails web application
(Section~\ref{sec:testhub}), which then displays the results to the
developers on a dynamic website.

\subsection{MESA}
\label{sec:mesa}

High-level \MESA\footnote{Source code available at 
\url{https://github.com/MESAHub/mesa}.}\ modules (currently \code{star}, \code{binary}, and
\code{astero}) each contain a \texttt{test\_suite} directory with a
collection of test cases.
\MESA\ includes a set of shell scripts that perform bookkeeping tasks such as
enumerating the available tests and providing a mapping between an
integer identifier and a given test.

A shell script \texttt{each\_test\_run} is responsible for running a
single test case and recording its output.  When invoked, it creates a
YAML\footnote{YAML is a human friendly data serialization
  standard: \url{https://yaml.org/}} file \code{testhub.yml} and populates it with metadata such as
the test name and the values of testing-related options.  The test is
executed by invoking shell scripts common to each test case directory that
compile (\texttt{mk}), run (\texttt{rn}), and restart (\texttt{re}) the test.  The run and restart scripts that
invoke \MESA\ are instrumented to record their memory usage.  The
output (to stdout and stderr) of these scripts is captured and in the
case of failing tests, transmitted by \mesatest\ to a remote server
for diagnostic purposes (Section~\ref{sec:testhub-logs}).

\begin{figure*}
  \includegraphics[width=0.5\textwidth]{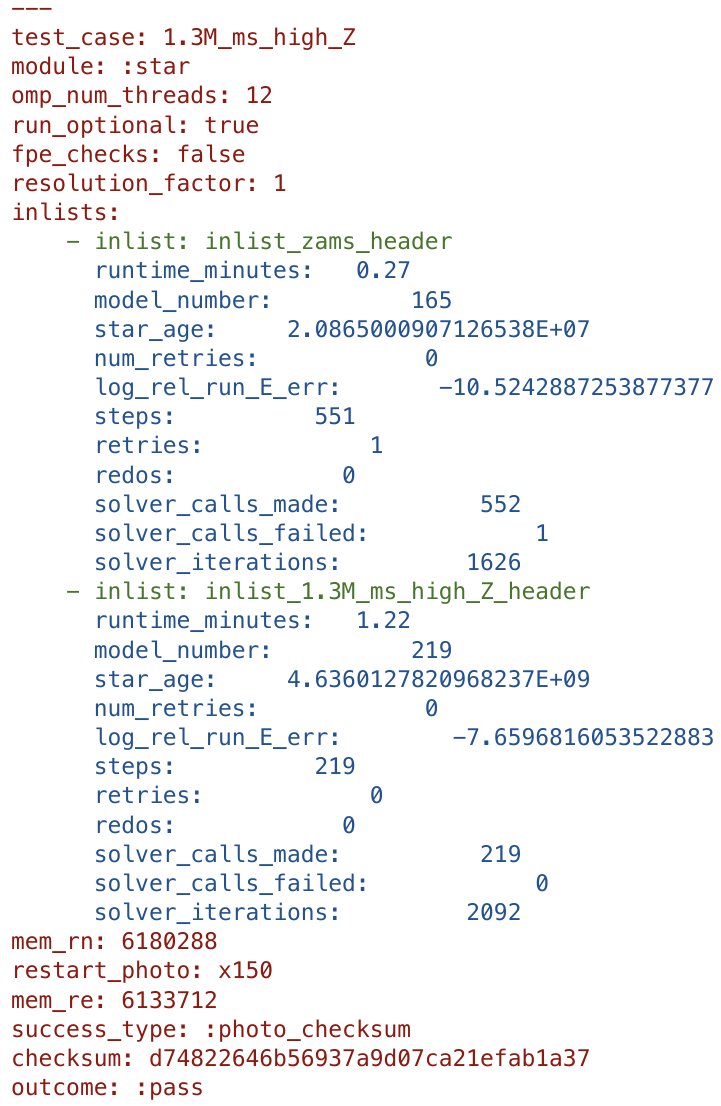}
  \centering
  \caption{Anatomy of a \texttt{testhub.yml} YAML file. Red text comes
    from \texttt{each\_test\_run}, green text from \texttt{do\_one},
    and blue text from \MESA\ itself.}
    \label{fig:yaml}
\end{figure*}

Within a test case, the \code{rn} script is responsible for executing the
test.  A set of helper scripts abstracts the process of running a
single \MESA\ inlist\footnote{An ``inlist'' is a generic name for a \MESA{} input
file which is a collection of Fortran namelists.  The main inlist may direct
\MESA{} to read other inlist files.} into a function called
\code{do\_one}.  A \texttt{rn} script is effectively a sequence of
\code{do\_one} commands.
Each call to \code{do\_one} records the name of the inlist file and
invokes the relevant \MESA{} executable.  When \MESA\ exits,
testing-specific code inserted into its \texttt{extras\_after\_evolve}
hook appends summary information about \MESA's performance such as the
runtime, number of steps, number of failed solver calls, and a SHA-256 hash
generated from the final model to the YAML file. Individual test cases
may optionally report physical quantities to the YAML file for longitudinal
tracking.
In this way, when the run is concluded, a YAML file summarizing the
complete run has been generated (see Figure~\ref{fig:yaml}).

What physical quantities to track is an area of ongoing work. Scalar quantities such as the 
final star mass, final core mass, or the maximum value of some quantity are trivial to track.
However, more complex quantities such as the time variation in the star's mass or the 
radial distribution of a quantity inside a stellar model cannot currently be tracked,
though we have had instances where this information would have caught a bug in \MESA.
In principle, the tests themselves could check time-varying quantities
or radial distributions and report out a failure if something is awry. In
practice, knowing what the ``correct'' configurations are and how to identify
meaningful deviations are difficult challenges. In a few instances, we have
instructed tests to output a png file with plots of relevant quantities for more
rapid diagnosis after a reported failure. However, such checks must still be
done by hand as the files themselves cannot yet be used to automatically detect
a failure, nor are they available online.

Many test cases allow for one or more of the slower \code{do\_one} steps to be
skipped if a particular environment variable is set. We refer to the inlists
associated with these skipped steps as ``optional'' inlists. Skipping optional
inlists is done to cut down on excessive runtimes, since a full run of all test
cases on a single workstation can take over 24 hours otherwise. To allow the
steps after a skipped step to still run, we keep saved models for the subsequent
steps to load.

Skipping optional inlists obviously exposes us to missed regressions, so we
have a cluster set up to do up to one full run (including optional inlists) on
the \code{main} branch each day. We have also devised a way to request a full
run on any commit (see section \ref{sec:commit_messages}). For a full run,
\code{testhub.yml} reports data for all inlists and indicates that it ran
optional inlists with a boolean flag, as in the top portion of 
Figure~\ref{fig:yaml}.

\subsection{mesa\_test}
\label{sec:mesa_test}
A Ruby gem \mesatest\footnote{Source code available at
\url{https://github.com/MESAHub/mesa_test}.}\ is responsible for obtaining a given version of
\MESA{} and using the built-in infrastructure described in the previous
subsection to compile \MESA\ and run its tests.  It then processes the
generated YAML files and transmits that information to the \MESA{} TestHub.

When \mesatest\ is first set up, it creates a git mirror repository on
the local machine.  This is a bare copy of the remote repository (hosted on
GitHub), meaning there is no associated working tree.  
When a user requests to test a specific version, the
mirror is updated (via a \code{git fetch})
and a working copy is created (via \code{git worktree}).  This process
ensures that a fresh working tree is created each time and avoids
having to repeatedly clone the multi-GB repository.

After \MESA\ is installed, tests can then be run.  The design of
\mesatest\ allows tests to be invoked individually, enabling
easy parallelization via schedulers (e.g., SLURM) in cluster computing
environments.  Individual \MESA\ developers have created testing
scripts tailored to the computing clusters to which they have access.

Figure~\ref{fig:flowchart} illustrates that \mesatest{} serves as the
messenger between the utilitarian shell scripts that control testing in \mesa
{} and the storage and dissemination role of the \MESA{} TestHub.  Keeping the role of \mesatest{} narrow and its code as simple as possible has proved beneficial.  Relying on testing scripts included within \mesa{} rather than reimplementing them in \mesatest{} reduces code duplication and removes opportunities for errors.  More importantly,
changes in the \MESA{} TestHub or in \mesa{} itself do not require any
software updates to occur on the testing computers.

With this separation of responsibilities, \mesatest{} does not need to be
updated often by developers or testers.  A common type of desired change is 
tracking a new piece of data for each test case.  To do so, the test cases simply write
out the new quantity to the YAML file, and a 
simple migration to the TestHub database allows it to accept those new quantities.
Meanwhile, \mesatest{} continues to pass the data from the YAML files to the
TestHub, agnostic to its contents.

\begin{figure}
\begin{algorithm}[H]
  \caption{Pseudo-code for the execution of \mesatest{} in parallel on a computing cluster}
  \label{alg:cluster}
  \begin{algorithmic}
    \Loop \ Every $X$ minutes
      \For{$commit$ in last $N$ commits}
        \If{$commit$ is untested} \par
        \hskip\algorithmicindent \Call{test\_commit}{$commit$}
        \EndIf
      \EndFor
    \EndLoop

    \Function{test\_commit}{$commit$} \par
      \hskip\algorithmicindent \Call{download\_and\_build\_mesa}{$commit$} \par
      \hskip\algorithmicindent \If{\Call{build\_succesful}{$commit$}}
        \For{$module$ in star binary astero}
          \ForAll{$tests$ in $module$} \par
          \hskip\algorithmicindent \Call{run\_test}{$commit$, $module$, $test$} \par
          \hskip\algorithmicindent \Call{submit\_result}{$commit$, $module$, $test$}
          \EndFor
        \EndFor
      \hskip\algorithmicindent \EndIf \par
      \Call{submit\_cleanup\_job}{$commit$}

    \EndFunction
    \end{algorithmic}
\end{algorithm}
\end{figure}

Algorithm~\ref{alg:cluster} shows the pseudo-code for how \mesatest{} is run on computing clusters, which we use to minimize the delay between committing code and seeing the results. Every $X$ minutes (usually in the 5--10 minute range), a cluster will start a management script via cron or scron that starts \mesatest{}. This script will then check for any new commits, since the last time it ran, and if there are new commits it will submit to the cluster's queue system a job that then runs \mesatest{} to download and build \MESA. While \mesatest{} can run the individual test cases serially, to maximize performance on a cluster we must run them in parallel. This is achieved by iterating over all test cases and submitting one job to the cluster's queue per test case. This job is then responsible for running the test, submitting the results to the TestHub, and potentially cleaning up any files. When running multiple \MESA's each with around 100 test cases, we can generate significant network bandwidth and use significant storage temporarily during a test. Combined with the fact that each computing cluster usually runs a different queueing system, this means we must optimize and refine the testing scripts for each new environment. A version of these scripts optimized for the Helios cluster at the University of Amsterdam can be found at \url{https://github.com/rjfarmer/mesa-helios-test}.

While an interface that needs to continuously poll GitHub for changes introduces a delay between when it can start a job, it is much simpler to implement on a shared computing cluster instead of an interface based on receiving push notifications from GitHub (like the TestHub website does).

\subsection{MESA TestHub}
\label{sec:testhub}
The \MESA{} TestHub\footnote{Source code available at
\url{https://github.com/MESAHub/MESATestHub}.} is a web app built on top of
Ruby on Rails that collects,
displays, and disseminates the result of testing conducted by computers
running \mesatest{}. Currently, it is hosted at
\url{https://testhub.mesastar.org}. It uses a PostgreSQL database to store data associated with the \mesa{}
repository as well as information about registered testers, their computers used for testing, and
their test result submissions.

\subsubsection{Syncing with GitHub} 
\label{ssub:syncing_with_github}

To organize results on a commit-by-commit basis, the TestHub must have
knowledge of the \mesa{} repository. To that end, the TestHub tracks all open
branches and their respective commits. To accomplish this, the \mesa{} repository on GitHub uses a
webhook that makes a request to the TestHub every time a developer pushes commits to GitHub.
Upon receiving this webhook request, the TestHub uses the GitHub
API to retrieve an updated listing of all open branches and their head
commits. If any branch is missing or out of date, it uses the GitHub API to
retrieve commit data from GitHub and update the TestHub's database.

\subsubsection{Interface with \mesatest{}} 
\label{ssub:interface_with_mesatest}
The TestHub implements a simple API where authenticated https requests can
deliver JSON payloads of test data which are then ingested into the database.
The \mesatest{} gem streamlines this process by taking the various YAML output
files from compilation and testing as well as computer information, converting 
this data to JSON, and then sending the result as a payload attached to an
authenticated request to the TestHub server using the API.

\subsubsection{The Web Front-End} 
\label{ssub:the_web_front_end}
To interact with the data it collects, the TestHub provides a rich front-end experience
on browsers that is largely powered by the free Bootstrap front-end framework
with some additional javascript features provided by jQuery. This is also
where registered users set up a computer in the database to receive testing
data. There are four primary useful views in the web interface.

\paragraph{Single Commit View} 
\label{par:single_commit}
\begin{figure*}
  \includegraphics[width=0.67\textwidth]{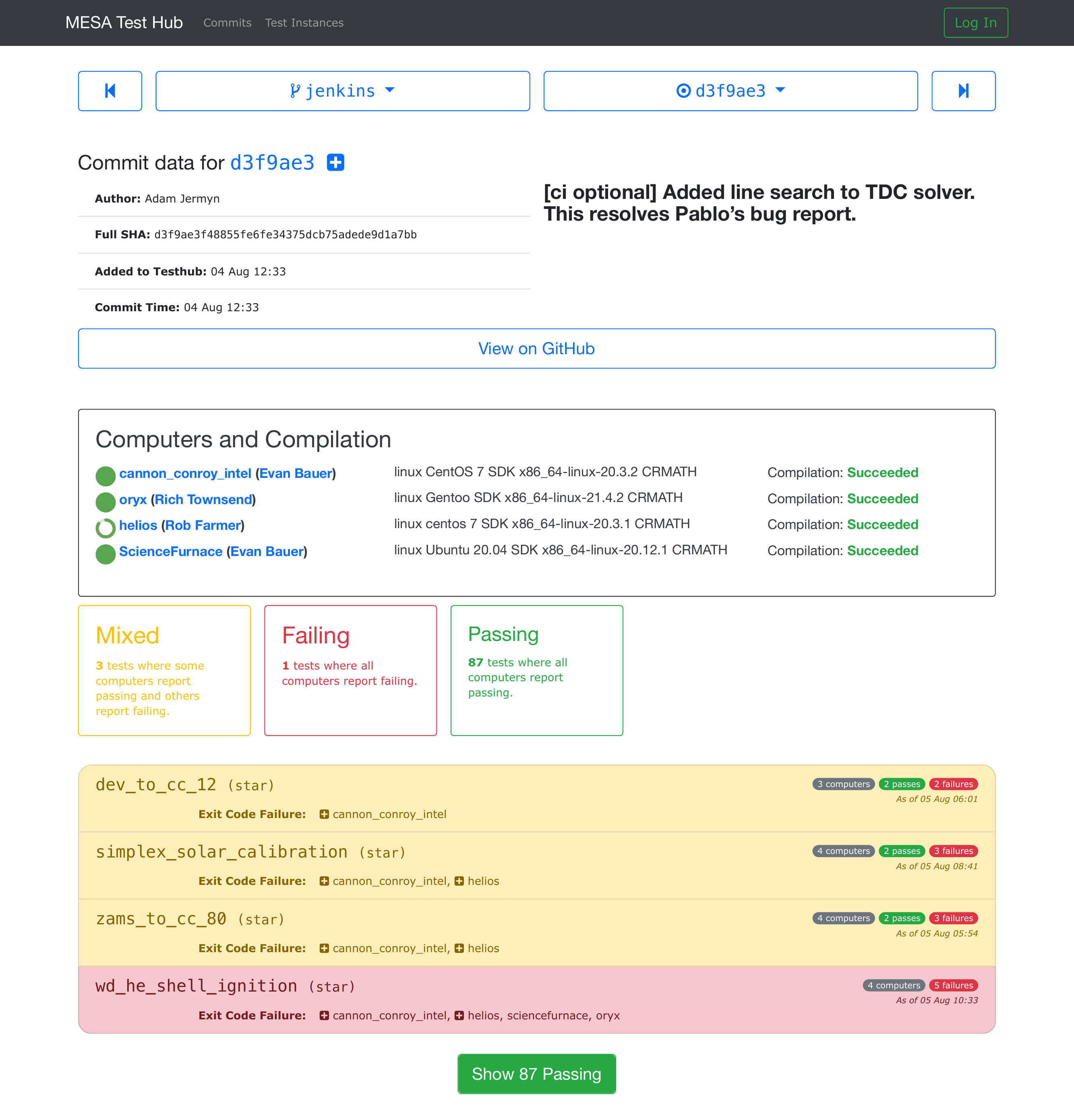}
  \centering
  \caption{An example of the single commit view. Basic information about the commit itself is shown at the top. Below that is a summary of the computers that have submitted tests, where full circles indicate all tests have been run, and an incomplete circle signifies that at least some test cases have not been submitted. Below that, we see three tests with a mixed status (yellow) and one that is failing on all test cases.}\label{fig:single_commit}
\end{figure*}
The single commit view shows typical data for a commit, including its full
SHA-256 hash, author, and message. It also shows a color-coded list of all
test cases present in that commit. Test cases can have a variety of statuses
depending on the results submitted so far. Tests can be \emph{passing} if
every instance of that test case submitted has passed and has either no
checksum or matching checksums (for multiple submissions). They can also be
\emph{failing}, where every instance has indicated a failure in the test. They
can be \emph{mixed} if there are at least one passing and one failing
instance.

Since most test case submissions include a checksum generated from the
final model, a test can also have a \emph{multiple checksum} status if all
instances pass, but different computers produce different checksums for the
final model. This would be concerning since most of \MESA{}
is written such that results should be bit-for-bit consistent across multiple
platforms \citep{Paxton2015}. Notably, we do not compare checksums between
submissions with optional inlists and
those without, since they would not be expected to match. Additionally, we do
not track checksums for several test cases that use code that is not yet
optimized to yield bit-for-bit consistency across multiple platforms, though
bringing these pieces of the codebase to bit-for-bit consistency is a long-term
goal of \MESA{} development.

Finally, test cases can be \emph{untested} if no instances have
been submitted yet. Figure~\ref{fig:single_commit} shows a typical instance of
this view.

\paragraph{Commits Index View} 
\label{par:commits_index}
\begin{figure*}
  \includegraphics[width=0.67\textwidth]{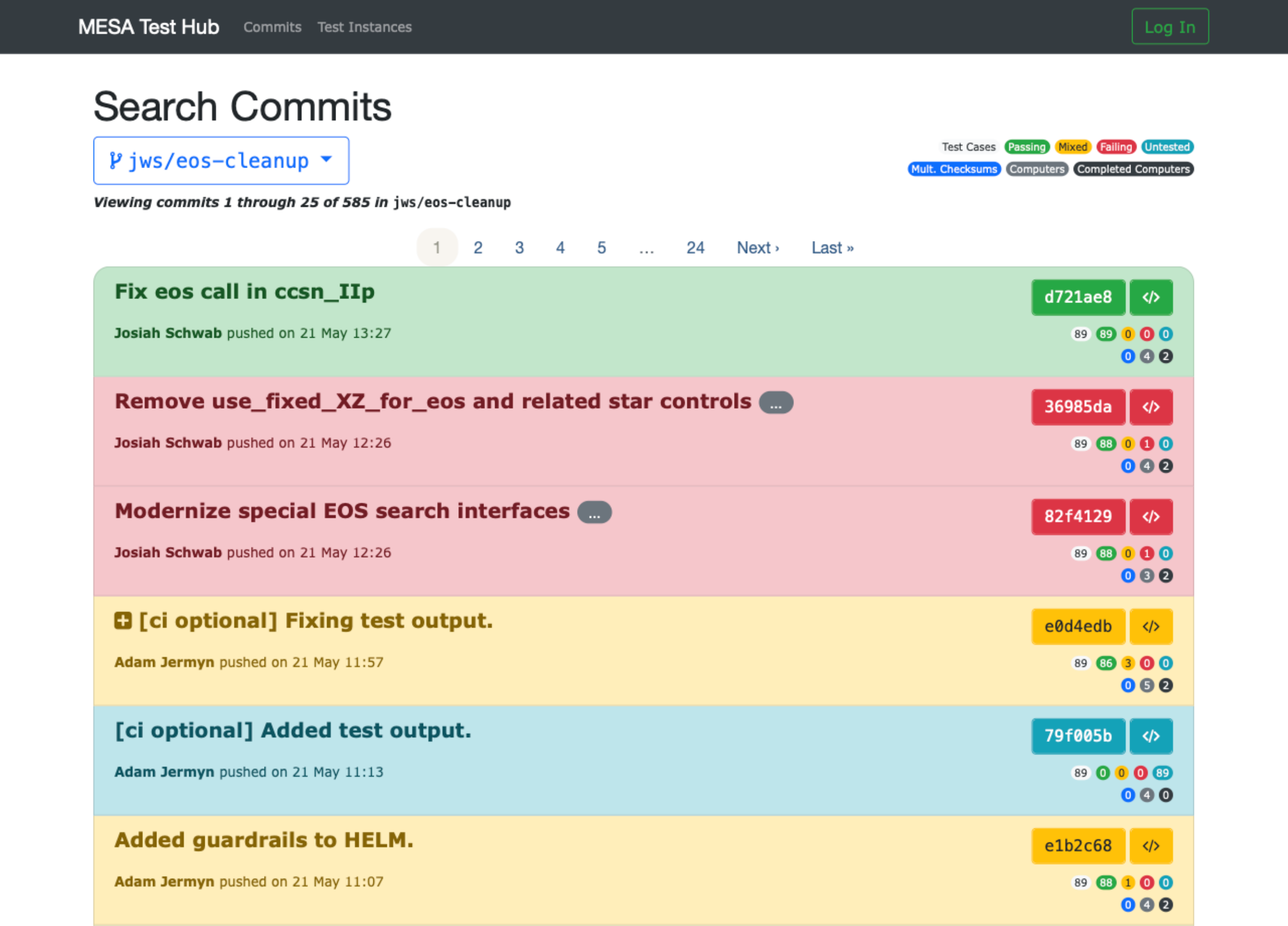}
  \centering
  \caption{An example of the commits index view showing commits that are passing, failing, mixed, and untested. The ``+'' icon indicates that all optional inlists were run for all tests on at least one machine for that commit.\label{fig:commits_index}}
\end{figure*}
The commits index view lists commits in a particular branch, indicating their
status in a color-coded list. Each entry shows basic information about the
commit and a collection of badges indicating the number of tests that pass,
fail, etc. The entire commit entry is color coded to reflect the status of
that commit's most alarming test case's status, giving a quick sense of how
that commit is faring. Figure~\ref{fig:commits_index} shows a typical example of this view.

In this view, one can also quickly toggle between different branches to see
commits in context of their recent development. This is particularly useful
when deciding if and when to merge one branch into another, since one wouldn't
want to merge in changes from another branch that are known to cause test
cases to fail.

\paragraph{Test Case Commit View} 
\label{par:test_case_commit_view}
\begin{figure*}
  \includegraphics[width=0.67\textwidth]{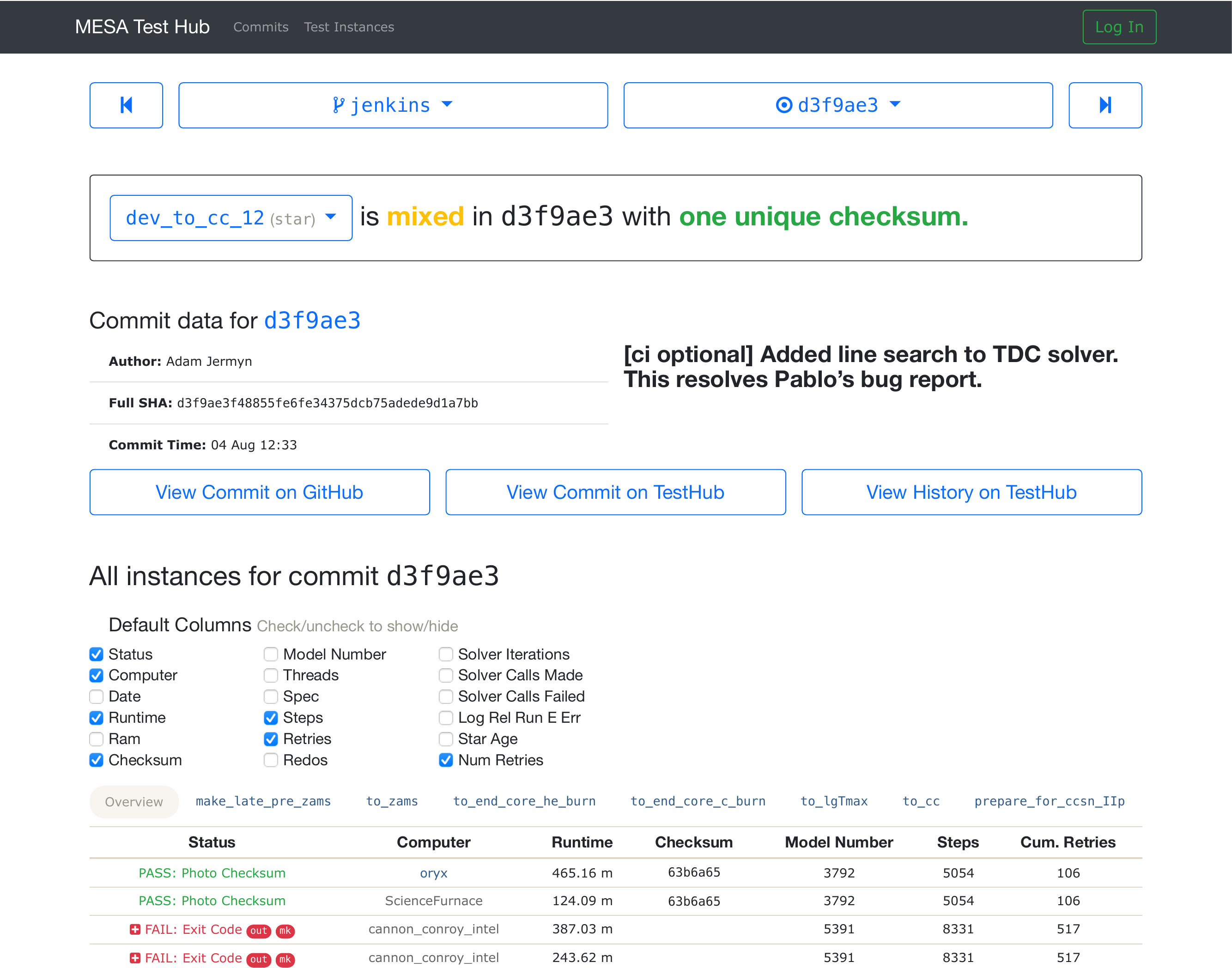}
  \centering
  \caption{An example of the test case commit view showing that the failing instances are coming from a single computer, along with links to view the logs of those instances (see section \ref{sec:testhub-logs}). The ``+'' icons indicate submissions where optional inlists were run. In this example, we can see that enabling optional inlists revealed a regression in this test.\label{fig:test_case_commit}}
\end{figure*}
The test case commit view shows data for a given test case on a given commit.
This shows why a test case has the indicated status in the single commit view by
listing all instances of that test case that have been submitted, including
granular data like checksum hashes, runtime, RAM usage, and physical data
related to that particular test case. Figure~\ref{fig:test_case_commit} shows an example of this view.

\paragraph{Test Case History View} 
\label{par:test_case_history_view}
\begin{figure*}
  \includegraphics[width=0.67\textwidth]{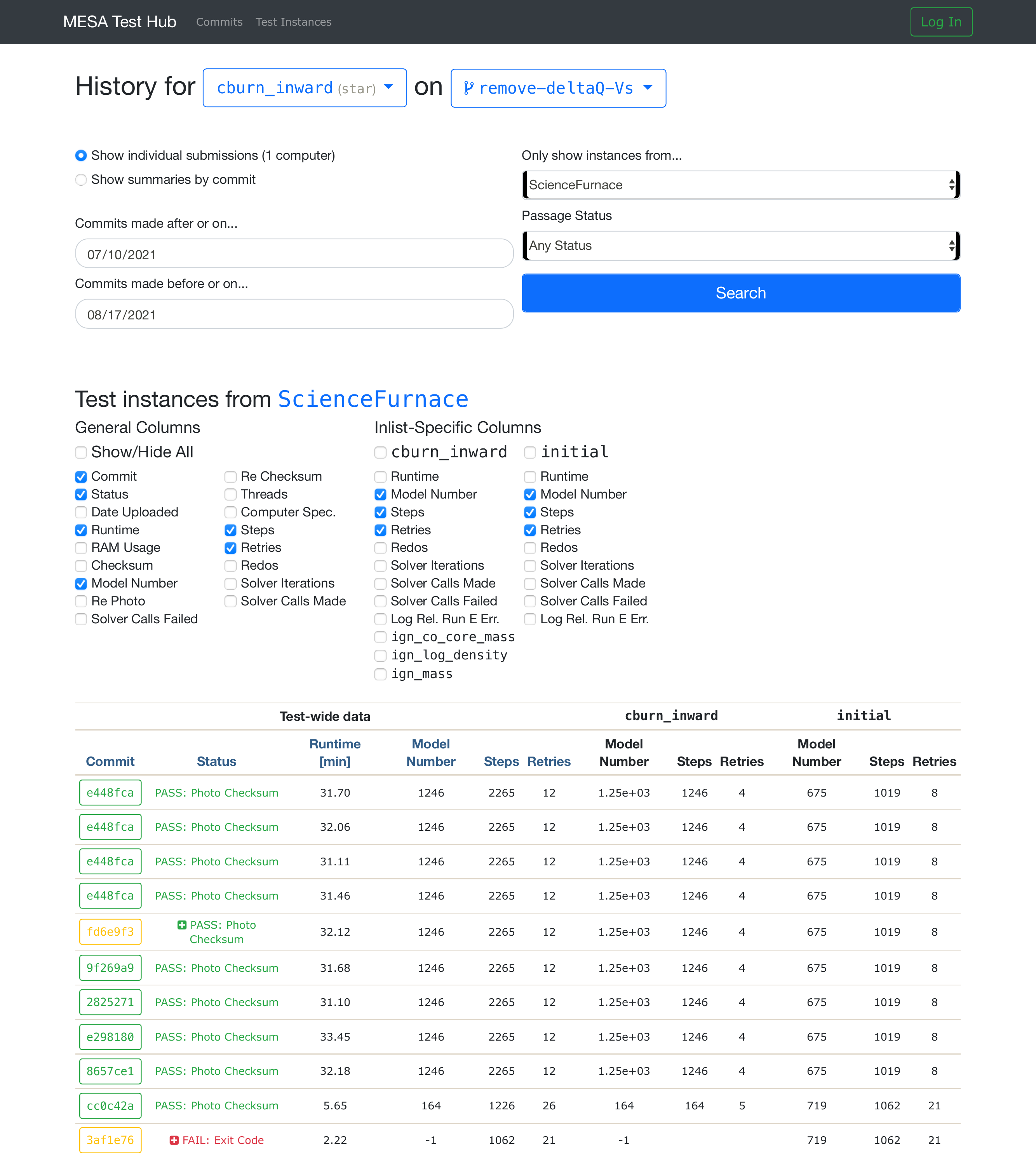}
  \centering
  \caption{An example of the test case history view showing all submissions from the computer \texttt{ScienceFurnace} for the test case \texttt{cburn\_inward} over a range of dates for all commits on the \texttt{remove-deltaQ-Vs} branch.\label{fig:test_case_history}}
\end{figure*}
The test case history view shows data for a given test case over many
commits, either by showing commit-wide summary data for each commit 
(essentially the data that is expected to agree among different computers on a
given commit) or by showing data from a particular submitting computer. This
is useful for seeing when a test case first started failing or how some
physical quantity varies with time (reflecting changes in the software). When looking at individual computers'
data, this view reveals trends in runtime and RAM usage over many commits. Figure~\ref{fig:test_case_history} shows an example of a computer-specific test case history view.

\subsubsection{Daily E-mail} 
\label{ssub:daily_e_mail}
\begin{figure*}
  \includegraphics[width=0.67\textwidth]{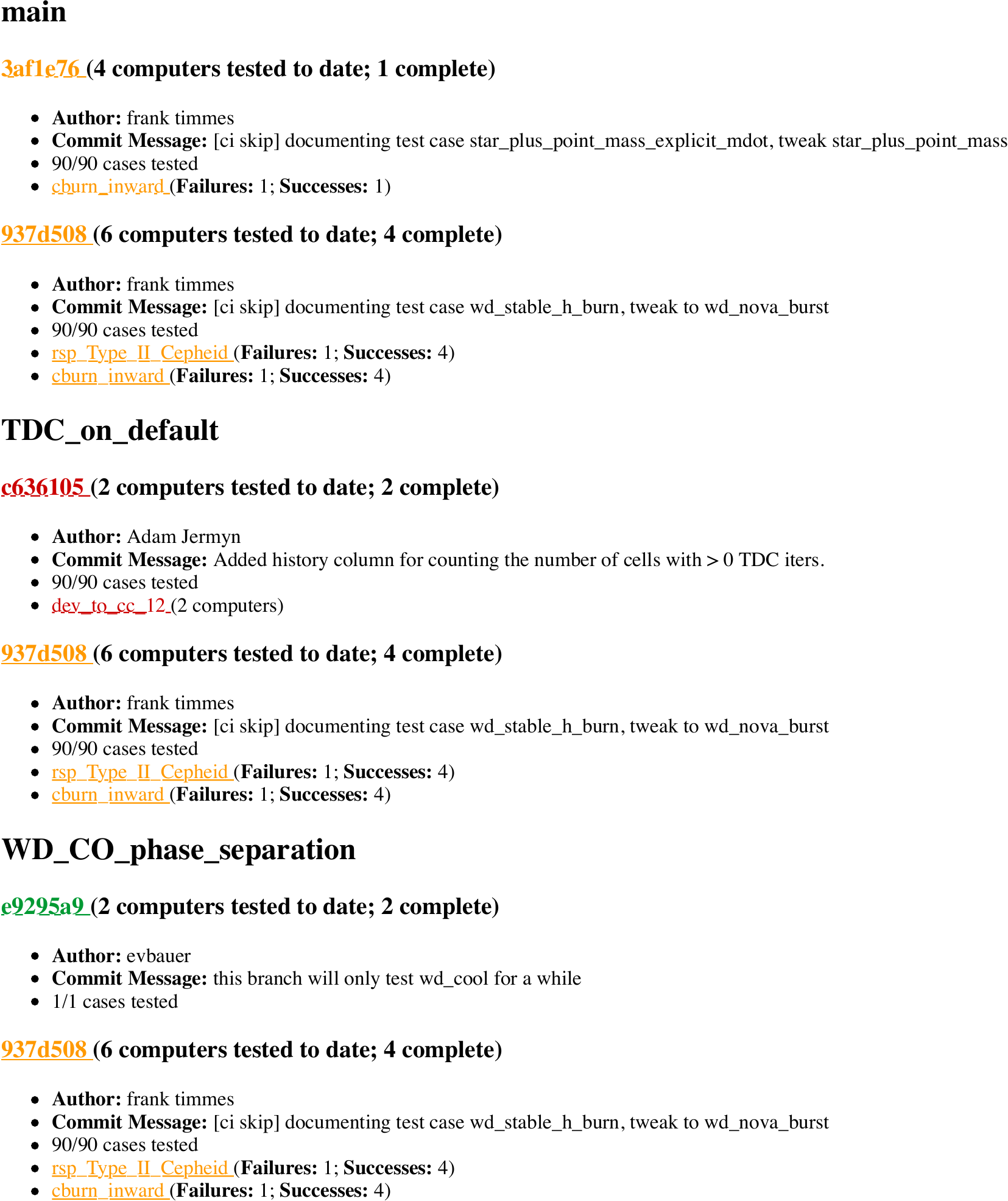}
  \centering
  \caption{An example of automatic daily summary email developers receive from the TestHub.\label{fig:daily_email}}
\end{figure*}
Being able to track the status of commits on demand via the web front-end is
invaluable, but since its inception the \MESA{} TestHub has implemented a
daily e-mail summarizing recent testing activity. Every commit in every branch
tested in the previous 24 hours is reported using a similar color scheme to
that indicated above, including a listing of all test cases that are failing,
mixed, or show multiple checksums with links to each relevant single commit 
view and test case commit view. The daily e-mail serves as a way to keep
the development team informed on the status of various branches. We have
also set these e-mails to be sent to a dedicated testing channel in the
development team's Slack workspace, where they often serve as the anchor
for debugging discussions. Figure~\ref{fig:daily_email} shows an example of
one such e-mail. In this case, the example e-mail shows the status of
ongoing development on branches that have since been merged into \code{main} to
incorporate major new features into recent MESA releases:
time-dependent-convection (TDC, \citealt{Jermyn2023}) and
carbon-oxygen phase separation \citep{Bauer2023}.

\subsubsection{Failure Logs}
\label{sec:testhub-logs}

Test cases inevitably fail.  When they do, it is important that
developers have easy access to the output associated with the failing
tests.  This is especially useful in the case of failures that are not
easy for an individual developer to reproduce (for example, because
the failure is intermittent or because it only occurs with a specific
operating system or compiler).  Even in the case of easily
reproducible failures, this saves a developer the time needed to
re-run the case and trigger the failure themselves, or to wait for the developer who
owns the machine to report the failure(s) manually (especially important given
the multiple time zones that \MESA{} development occurs over).

When a compilation or test failure occurs, \mesatest\ transmits the information
that was output (i.e., the contents of stdout and stderr, not the data written
to the \MESA{} history or profile files) to a separate log archiving server that
is tightly integrated with the TestHub. Transmitting only the failing cases
saves significant bandwidth and storage as the full logs from a complete test
suite run are $\gtrsim 10$ MB in size.  Because these files are intended for
diagnosing failures, they are only retained for a limited time, under the
assumption their utility is largely exhausted after issues are fixed. On the
TestHub web interface, notifications of failing compilations and failing test
case instances are accompanied by links to the relevant log files if they
exist.

The \mesatest\ gem also includes an option to force submission of
test logs even for cases where no failure occurred. We currently have
one cluster set up to force test log submissions to the latest commit
on the {\tt main} branch of MESA up to once per day. This allows
developers to have a quick reference to test output for passing test
cases on {\tt main} when tests report failures on other branches,
while also avoiding an overload of test log files.

\subsubsection{Hosting and Hardware} 
\label{ssub:hosting_and_hardware}
The TestHub is currently hosted by the Heroku\footnote{https://heroku.com} platform as a service on a single ``basic'' dyno. Dynos are are isolated, virtualized Unix containers designed to work with a particular application environment. In the TestHub's case, the dyno is tuned to work well with Ruby on Rails apps. The basic level is the least capable of Heroku's persistent plans, providing 512 MB of RAM and modest access to a CPU (Heroku does not publish precise details on what CPUs or how many cores a given dyno has access to) at a cost of \$7 USD per month as of August 2023. New deployments are triggered by issuing a \texttt{git push} to a special \texttt{heroku} remote, with the actual build deployment handled automatically on Heroku's end. The runtime environment can be accessed with command line access through Heroku's website or via their command line tool.

In addition to the web server, Heroku also hosts the PostgreSQL database the TestHub relies on via their Heroku Postgres add-on service. We use the ``basic'' level of service, which allows 10,000,000 rows, 10 GB of storage, and up to 20 simultaneous connections at a cost of \$9 USD per month as of August 2023. Over the first 32 months of usage since \mesa{} development shifted to GitHub, the TestHub now uses over 2,400,000 rows spread over 13 tables, consuming 564.5 MB. So in the medium term, we may require upgraded database service, or we may opt to purge old test data to keep the data storage lean.

With the relatively low throughput the TestHub experiences, it could be served by a rather inexpensive server and a reliable internet connection. However, the low cost of \$16 USD per month and reliable service has made Heroku's platform as a service a robust and simple hosting solution. Even at the modest levels of service the TestHub runs on, it does not appear to be meaningfully bottlenecked by the processor, RAM, or database connection.

\subsection{Commit messages} 
\label{sec:commit_messages}

To extend the flexibility in testing we have introduced a number of commit strings to pass data to the
testing machines. These strings, if present in the commit message, signal a request from the committer to the
testing machines to alter the standard testing setup in some way (note these are only requests as not every test machine respects every commit message):

\begin{itemize}

  \item $[\rm{ci\ skip}]$ This instructs testing machines not to run the test suite on this commit. This is primarily useful when committing changes that do not affect the code, for instance documentation changes. A basic check for passing compilation is still performed and submitted.
  \item $[\rm{ci\ optional}]$ Requests that testing machines run all optional inlists in all test cases. This is particularly useful right before
    attempting to merge a branch back into \code{main} to ensure no lurking
    regressions remain.
  \item $[\rm{ci\ fpe}]$ Requests that a number of additional compile time flags are enabled, as well as certain runtime flags that initialize data to NaN. This allows better testing for things like uninitialized variable usage and bit-for-bit issues.
  \item $[\rm{ci\ converge}]$ Request that MESA runs with an increased spatial and temporal resolution, to find test cases that are numerically unconverged.

\end{itemize}
In the past, we also had a [ci split] flag that signaled that the two
main clusters we were using at the time should split the tests. That is, each
cluster would only run half of the tests, but between the two, all tests would 
be run once. This allowed for quicker turnaround during periods of rapid
development. However, this required a lot of hand-tuning to balance which tests
went to which cluster and sacrificed the knowledge  that tests worked the same
on both systems (let alone that all tests could pass on a single system). Since
then, we've usually had enough available computing resources that the small
amount of time saved by this is not worth the complexity it introduced.

These flags enable greater flexibility in the testing. The committer can decide whether they wish for no testing ($[\rm{ci\ skip}]$), normal testing (no flag), or a slower but more in-depth testing ($[\rm{ci\ optional}]$ and/or $[\rm{ci\ converge}]$). By not having every machine respect all the flags we also ensure that the test suite is still run regularly. Indeed, we have several machines set up to test the latest commit on the main branch with all optional inlists and/or forcing the submission of logs to the log server. As a result, we will often have commits with the [ci skip] flag tested heavily if they happen to be the latest commit when this task is performed. Additional icons are also present in the web overview (See figures \ref{fig:single_commit}, \ref{fig:commits_index}, and \ref{fig:test_case_commit}) to show at a glance
which machines have tested the commit with each flag.

\section{Conclusions}
The combination of testing infrastructure inside \MESA{}, \mesatest{}, and the
\MESA{} TestHub has gone through many refinements as we have learned what works
well and what our development team really needs from a testing system.
Other continuous testing infrastructures exist, but the unique needs of the
\MESA{} project require a more bespoke solution. In particular, useful tests
in \MESA{} are long, complex stellar evolution problems and not simple unit
tests. We also value being able to report and track a
heterogenous set of metrics, including test-specific physical quantities.
Additionally, we need a testing framework designed to accommodate a wide
variety of systems rather than being focused on a single server in the cloud,
since \MESA{} users work at a variety of scales on multiple platforms.
And so, we designed a system that works well on individual laptops and computing
clusters; we have even adapted it to work with a continuous testing server at
the Flatiron Institute running Jenkins\footnote{https://jenkins.io}.

We conclude with some lessons learned over the development of \MESA's testing
infrastructure. First, separation of responsibilities is very important.
Originally, \mesatest{} handled downloading, installing, running of tests, collection of data, and
submission of results to the TestHub. This resulted in asking testers to
manually update their local version of \mesatest{} whenever a new piece of data
was to be collected or a bug needed to be fixed. Since moving the management of individual tests and test data to \MESA{} itself, \mesatest{} is much leaner and requires less
frequent updating and maintenance.

Secondly, balancing coverage of tests with speed requires thoughtful compromise.
More rigorous testing usually means a longer delay between a commit and a result,
as well as a larger investment of limited CPU hours. On the other hand, sparse
coverage runs the risk of missing important regressions. Our current
method of using optional inlists to skip the  most time-consuming portions of
tests has resulted in relatively fast turnarounds while still allowing for more
in-depth testing on at least a daily basis. However, we may need to make other
compromises such as dividing the test cases into ``probative'' and ``exemplary''
categories for prioritized testing if the test suite grows appreciably in size
without an accompanying growth in computational resources.

Thirdly, using a platform as a service (like Heroku) has been a smooth and
cost-effective experience. The portable nature of a cloud service was
particularly helpful as the lead developer of the TestHub switched institutions
with virtually no disturbance to the testing infrastructure.

Fourthly, while automated testing has caught many bugs and issues in \MESA, manually 
inspecting the output of the \MESA{} test suite is still needed from time to time. Certain
problems have arisen in the past that cannot be easily checked for by automated scripts.
Thus a human review of diagnostic plots and model output is still needed occasionally to 
ensure that \MESA{} is running correctly.
Automated testing is not meant to constitute a complete replacement
for the maintenance work of a team of experts covering the wide range
of physics encountered in stellar evolution, but rapid turnaround of
broad automated testing enabled by this platform has greatly expedited
the pace of development and the rate at which issues are discovered
and referred to experts for closer inspection.

Finally, we note that building and maintaining a robust testing infrastructure
for a project with complex needs like that of \MESA{} requires substantial
expertise and effort at multiple levels. In the case of the TestHub and
\mesatest, we needed some degree of full stack web development knowledge, but
the infrastructure within \MESA{} required a deep understanding of the way tests
can fail, how that information can be aggregated, and what information would be
essential for diagnosing regressions. As a research software project
grows from a single scientist to a distributed team, developers should be aware
of the testing and quality assurance needs of their project and not overlook the
significant investment of time and effort required to enable smooth development.

The \MESA{} testing infrastructure has been a critical factor in enabling the
continued development of this software. In the first two and a half years since
the move to git, we have averaged testing five commits per day, with
over 570,000 test case instances reported to the TestHub representing nearly a
million CPU hours of computation on clusters, workstations, and laptops. We hope
that by documenting our testing infrastructure, other projects in computational
science with similar demands might be able to build a system that supports their
further development.

\begin{acknowledgments}
  We thank the anonymous reviewer for their helpful comments and suggestions.
  W.W. acknowledges support by the National Science Foundation through grant AST-2238851.
  J.S. acknowledges support by NASA through Hubble Fellowship grant \# HST-HF2-51382.001-A awarded by the Space Telescope Science Institute, which is operated by the Association of Universities for Research in Astronomy, Inc., for NASA, under contract NAS5-26555, by the A.F. Morrison Fellowship in Lick Observatory, and by the National Science Foundation through grant ACI-1663688.  We acknowledge use of the lux supercomputer at UC Santa Cruz, funded by NSF MRI grant AST 1828315, and we thank Josh Sonstroem and Brant Robertson for their tireless efforts providing this resource.  We acknowledge use of the University of Amsterdam's Helios cluster which was supported by a European Research Council grant 715063, P.I S.E.~de~Mink.
  We acknowledge the use of the FASRC Cannon cluster supported by the
  FAS Division of Science Research Computing Group at Harvard
  University, with special thanks to Charlie Conroy and the Institute
  for Theory and Computation for access to their dedicated nodes on Cannon.
\end{acknowledgments}

\software{
\texttt{MESA} \citep[][\url{http://mesa.sourceforge.net}]{Paxton2011,Paxton2013,Paxton2015,Paxton2018,Paxton2019,Jermyn2023}, Ruby on Rails (\url{https://rubyonrails.org}), Bootstrap (\url{https://getbootstrap.com}), jQuery (\url{https://jquery.com})
}

\bibliography{testing}
\bibliographystyle{aasjournal}

\end{document}